\documentclass[preprint,11pt]{elsarticle}

\usepackage{fancyhdr}
\usepackage{fullpage}
\usepackage{amsmath}
\usepackage{amsbsy}
\usepackage{amssymb}
\usepackage{amscd}
\usepackage{amsfonts}
\usepackage{supertabular}
\usepackage{graphics}
\usepackage{verbatim}
\usepackage{epsfig}
\usepackage{xspace}
\usepackage{euscript}
\usepackage{alltt}
\usepackage{boxedminipage}
\usepackage{float}
\usepackage[colorlinks]{hyperref}
\usepackage{color}
\usepackage[all]{xy}
\usepackage{t1enc}
\usepackage{times,exscale}
\usepackage{graphicx,calc}
\usepackage{subfig}
\usepackage[ruled,vlined]{algorithm2e}
\usepackage{epstopdf}
\usepackage{array,multirow}

\usepackage{graphicx}					
\usepackage{amsmath,amssymb}		
\usepackage{subfig}					
\usepackage{multirow}
\usepackage{booktabs}

\usepackage{comment}

	\newcommand{\pp}{\mathrm{psp}}
	\newcommand{\ael}{\mathrm{ae}}
    
    	\newcommand{\bra}[1]{\langle#1|}
	\newcommand{\ket}[1]{|#1\rangle}
	\newcommand{\braket}[2]{\langle#1|#2\rangle}
	\newcommand{\enr}[3]{\langle#1|#2|#3\rangle}

\journal{arXiv}

\begin{document}

	\begin{frontmatter}
		
		\title{Soft and transferable pseudopotentials from multi-objective optimization}
		\author[gatech]{Mostafa Faghih Shojaei}
		\author[llnl]{John E. Pask}
		\author[gatech]{Andrew J. Medford}
		\author[gatech]{Phanish Suryanarayana\corref{cor}}
		\address[gatech]{College of Engineering, Georgia Institute of Technology, Atlanta, GA 30332, USA}
		\address[llnl]{Physics Division, Lawrence Livermore National Laboratory, Livermore, CA 94550, USA}
		\cortext[cor]{Corresponding Author (\it phanish.suryanarayana@ce.gatech.edu) }
		
		\begin{abstract}
Ab initio pseudopotentials are a linchpin of modern molecular and condensed matter electronic structure calculations. 
In this work, we employ multi-objective optimization to maximize pseudopotential softness while maintaining high accuracy and transferability.
To accomplish this, we develop a formulation in which softness and accuracy are simultaneously maximized, with accuracy determined by the ability to reproduce all-electron energy differences between Bravais lattice structures, whereupon the resulting Pareto frontier is scanned for the softest pseudopotential that provides the desired accuracy in established transferability tests. We employ an evolutionary algorithm to solve the multi-objective optimization problem and apply it to generate a comprehensive table of optimized norm-conserving Vanderbilt (ONCV) pseudopotentials (\url{https://github.com/SPARC-X/SPMS-psps}). 
We show that the resulting table is softer than existing tables of comparable accuracy, while more accurate than tables of comparable softness. 
The potentials thus afford the possibility to speed up calculations in a broad range of applications areas while maintaining high accuracy.
		\end{abstract}
		
		
	\end{frontmatter}

	\section{Introduction}
Over the past few decades, Kohn–Sham density functional theory (DFT) \cite{DFT1,DFT2_kohn1965self} has established itself as a cornerstone of physical, chemical, and materials research, enabling the study of a wide variety of systems from the first principles of quantum mechanics, with no empirical or adjustable parameters. The popularity of DFT can be attributed to its high accuracy-to-cost ratio relative to other such ab initio methods, particularly in the context of the widely adopted pseudopotential approximation \cite{Martin2004}. In this approach, core electrons not participating in the chemistry of interest are removed from the calculation by replacing the Coulomb potential of the nucleus by an effective potential corresponding to the nucleus and frozen core electrons, referred to as a pseudopotential. The goal of such a strategy is to significantly reduce the computational cost of the  calculations while keeping the physical/chemical properties of the system sufficiently unchanged. 

The pseudopotential formulations most widely employed in modern Kohn-Sham calculations are of three main kinds: ultrasoft \cite{USPPs}; projector-augmented-wave (PAW) \cite{PAW_PSPs}; and norm-conserving, such as Hamann-Schl\"{u}ter-Chiang (HSC) \cite{NCPSPs_first}, Kerker \cite{kerker1980non}, Bachelet-Hamann-Schl\"{u}ter (BHS) \cite{NCPSPs_set}, generalized norm-conserving \cite{GNPPs}, Rappe-Rabe-Kaxiras-Joannopoulos (RRKJ) \cite{RRKJ_optimization}, Troullier-Martins (TM) \cite{TM_psps}, Goedecker-Teter-Hutter (GTH) \cite{GTH_anaytical}, Hartwigsen-Goedecker-Hutter (HGH) \cite{HGH}, and Hamann's optimized norm-conserving Vanderbilt (ONCV) \cite{ONCVPSP} potentials. Among these, ultrasoft and PAW  are typically softer than norm-conserving potentials, i.e., a larger grid spacing in real-space calculations or smaller energy cutoff in planewave calculations suffices to achieve a specified accuracy with respect to discretization. However, ultrasoft and PAW formulations lead to a generalized rather than standard eigenvalue problem, even within orthonormal discretization schemes. This can increase computational cost and limit parallel scalability, in real-space methods \cite{beck2000rsmeth, saad2010esmeth, SPARC} in particular, 
where efficient and scalable preconditioners for the eigenvalue problem are lacking. In addition, the PAW formulation introduces complexities in deriving and implementing expressions involving derivatives, such as atomic forces, stress tensor, and phonons. These and other issues have motivated the further development of norm-conserving pseudopotentials in recent works \cite{ONCVPSP,SG15,PseudoDojo}.

The generation of soft and transferable norm-conserving pseudopotentials can, however, be a challenging task. To simplify the process of performing DFT simulations, a number of pseudopotential tables, i.e., sets of pseudopotentials covering most of the periodic table, have been generated. These include (i) ultrasoft: Garrity-Bennett-Rabe-Vanderbilt (GBRV) \cite{GBRV}; (ii) PAW: VASP \cite{kresse1999ultrasoft}, Jollet-Torrent-Holzwarth (JTH) \cite{Delta_prime_2014}, Topsakal-Wentzcovitch \cite{TOPSAKAL2014263}; (iii) norm-conserving: BHS \cite{NCPSPs_set}, GTH \cite{GTH_anaytical}, HGH \cite{GTH_database_2}, PARSEC \cite{PARSECwebPSP}, Fritz-Haber-Institute (FHI) \cite{fuchs1999ab}, Krack \cite{GTH_database_2}, Willand et al. 
\cite{willand2013norm}, Schlipf-Gygi (SG15) \cite{SG15}, and PseudoDojo \cite{PseudoDojo}; and (iv) mixtures of these types: pslibrary \cite{pslibrary_accuracy,pslibrary} and SSSP \cite{SSSP}. However, though these pseudopotentials have been tested for their accuracy, apart from SG15 \cite{SG15}, where a metric including both accuracy (in terms of lattice constant error) and softness was maximized, the pseudopotentials have not been generated through a systematic optimization process. This makes the generation of pseudopotential tables an arduous and time consuming task, particularly given the large number of  variants of interest in practice: exchange-correlation functionals, relativistic effects, core-valence partitions, and the like.
In addition, it is likely that these pseudopotentials are harder than necessary for the accuracy they provide.
 

In this work, we employ multi-objective optimization to maximize pseudopotential softness while maintaining high accuracy and transferability.
To accomplish this, we develop a formulation in which softness and accuracy are simultaneously maximized, with accuracy determined by the ability to reproduce all-electron energy differences between Bravais lattice structures, whereupon the resulting Pareto frontier is scanned for the softest pseudopotential that provides the desired accuracy in established transferability tests. We employ the optimization scheme to generate a table of ONCV pseudopotentials for sixty-nine chemical elements (H--La and Hf--Bi) within the Perdew–Burke–Ernzerhof (PBE) \cite{perdew1996generalized} exchange-correlation approximation (\url{https://github.com/SPARC-X/SPMS-psps}).  
We show that the resulting table is softer than recent tables of comparable accuracy, while more accurate than recent tables of comparable softness.

The remainder of this paper is organized as follows. In Section~\ref{formulation}, we develop a formulation for generating soft and transferable pseudopotentials. In Section~\ref{oncvpsp}, we provide an overview of the ONCV pseudopotential formalism. In Section~\ref{Sec:Implementation}, we describe the implementation for generating soft and transferable ONCV pseudopotentials. In Section~\ref{results}, we discuss some representative results as well as the accuracy and softness of the table of ONCV pseudopotentials generated. Finally, we provide concluding remarks in Section~\ref{conclusions}.

	\section{Formulation} \label{formulation}

We now present an approach for  generating soft and transferable pseudopotentials within any given pseudopotential formalism.  In particular, we consider  the following multi-objective optimization problem:
\begin{equation} \label{Eq:MultiObj}
\mathbf{p}^{*} = \arg \min_{\mathbf{p} \in P} \left(\mathcal{E}(\mathbf{p}), \frac{1}{\mathcal{S}(\mathbf{p})} \right) \,,
\end{equation}
where $\mathbf{p}$ is the vector of parameters that characterize the pseudopotential, $P$ is the feasible set of all such vectors, and $\mathcal{E}$ and $\mathcal{S}$ are metrics that measure the error and softness of the pseudopotential, respectively. Note that these metrics measure distances with respect to the all-electron Coloumb potential, and have been written as a function of  $\mathbf{p}$ to indicate the dependence of the calculated values on $\mathbf{p}$. Since $\mathcal{E}$ and  $1/\mathcal{S}$ have an overall negative correlation with each other, the solution of the above minimization problem is a Pareto set/frontier, i.e., for any pseudopotential that does not belong to the Pareto frontier, there exists at least one pseudopotential in the Pareto frontier that has smaller values for both $\mathcal{E}$ and $1/\mathcal{S}$. 

A natural variable for defining the softness metric $\mathcal{S}$ is the grid spacing used in real-space calculations,  or equivalently, the inverse of the energy cutoff used in planewave calculations.  However, there is no such  computationally tractable  \emph{universal}  error metric $\mathcal{E}$, i.e., one that captures the accuracy of the pseudopotential for every possible electronic environment  encountered in molecular and condensed matter systems. Indeed, if such a \emph{universal} $\mathcal{E}$ were to be available, then the softest pseudopotential for the desired accuracy can immediately be chosen from the generated Pareto frontier. In view of this, we reformulate the optimization problem in Eq.~\ref{Eq:MultiObj} as the following two-step scheme:
\begin{subequations}
\begin{eqnarray}
\hat{\bf p}  & = & \arg \min_{\mathbf{p} \in P} \left(\mathcal{E}_0(\mathbf{p}),\frac{1}{\mathcal{S}(\mathbf{p})} \right)  \,,  \label{Eq:Eq:MultiObjRef} \\
\mathbf{p}^*  & = &  \arg \min_{\mathbf{p} \in P_f} \frac{1}{\mathcal{S}(\mathbf{p})} \,, \quad  P_f = \{\hat{\bf p}: \mathcal{E}_i(\hat{\bf p}) \leq \epsilon_i, i=1, 2, \ldots, n \} \,, \label{Eq:Eq:MultiObjRefC} 
\end{eqnarray}
\end{subequations}
where $\mathcal{E}_0$ is some relatively simple and computationally efficient error metric that is used  in the multi-objective optimization,  and $\{\mathcal{E}_i\}_{i=1}^n$ are the more sophisticated and likely more computationally expensive  error metrics to be used in  choosing the pseudopotential from the Pareto frontier, with $\{\epsilon_i \}_{i=1}^n$ being the desired upper bounds on these errors. Within this reformulation, the error metric $\mathcal{E}_0$ is expected to reliably  estimate the accuracy and transferability of the pseudopotential, while simultaneously ensuring that the computational cost associated with the optimization is tractable. A maximally soft pseudopotential from the Pareto frontier with the desired accuracy and transferability for the physical/chemical application of interest can then be selected by suitably choosing $\{\mathcal{E}_i\}_{i=1}^n$ and $\{\epsilon\}_{i=1}^n$.

\begin{figure}[t]
		\centering
		\includegraphics[width = 0.85\textwidth]{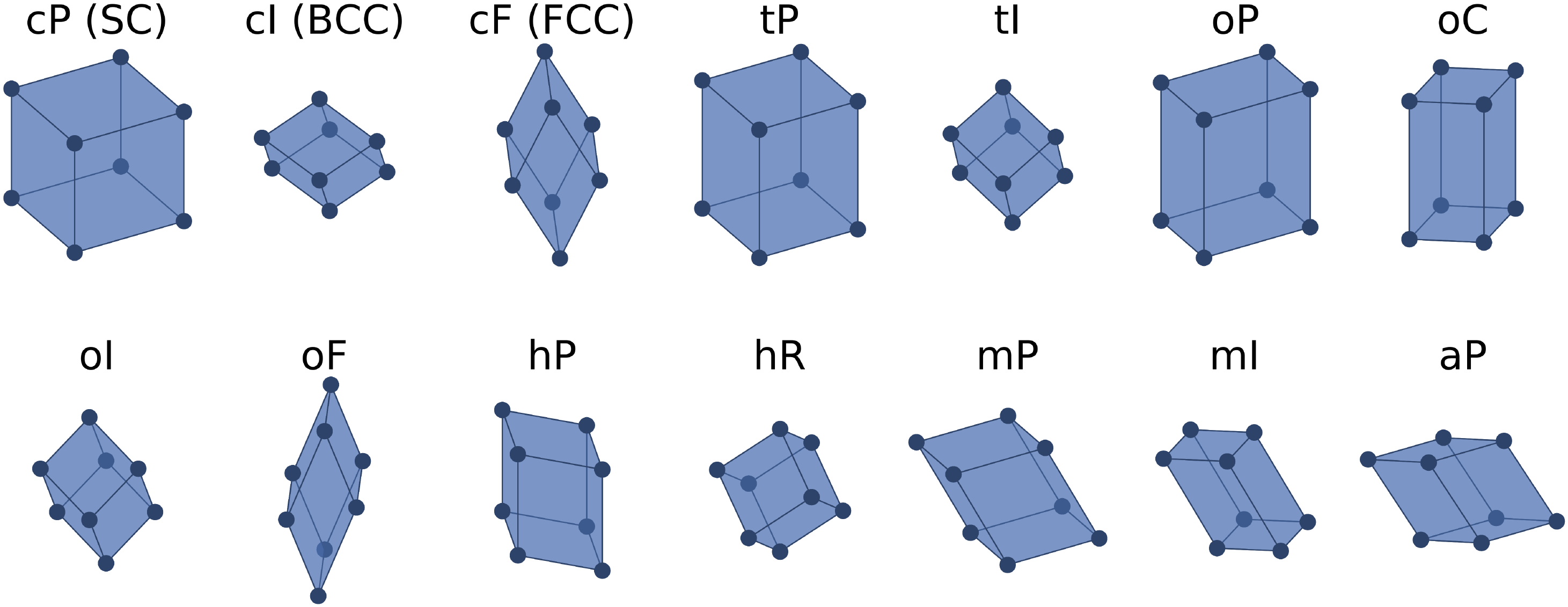} 
		\caption{The set of $14$ Bravais lattice symmetries used for defining the error metric $\delta_{\Delta E}$ in the multi-objective optimization (Eq.~\ref{Eq:Eq:MultiObjRef1}).}
		\label{cells}
	\end{figure}

We propose the use of a structural energy difference based error metric in the multi-objective optimization (Eq.~\ref{Eq:Eq:MultiObjRef}):
\begin{subequations}
\begin{eqnarray}
	\mathcal{E}_0 & := &	\delta_{\Delta E} = \frac{1}{\sqrt{14 m}}\left\lVert\Delta \boldsymbol{E}^\pp-\Delta \boldsymbol{E}^\ael\right\lVert_{2} \,, \label{Eq:Eq:deltadeltaE} \\
\Delta \boldsymbol{E}^{\pp/\mathrm{ae}} & = & (E^{\pp/\mathrm{ae}}_1-E^{\pp/\mathrm{ae}}_0, E^{\pp/\mathrm{ae}}_2-E^{\pp/\mathrm{ae}}_0, \ldots, E^{\pp/\mathrm{ae}}_{14 m}-E^{\pp/\mathrm{ae}}_0) \,,
\end{eqnarray}
\end{subequations}
where $\{E^\pp_i\}_{i=1}^{14m}$ and $\{E_i^{\rm ae}\}_{i=1}^{14m}$ are the Kohn-Sham energies for  a set of $14m$ primitive Bravais lattices: containing structures with the 14 Bravais lattice symmetries (Fig.~\ref{cells}) with $m$ different nearest neighbor distances. In addition, $E^{\pp/\mathrm{ae}}_0$ is a reference energy to ensure meaningful comparison between pseudopotential and all-electron results. Such a choice for $\mathcal{E}_0$ can be interpreted as follows: the primitive Bravais lattices represent a basis for sampling the various electronic environments that are encountered within materials systems. Indeed, the basis can be made more complete by considering Bravais lattices with a basis that have two or more different chemical elements. However, this comes with significant additional computational cost and complexity, and hence not considered in this work. For the softness metric, we choose the following:
\begin{equation}
\mathcal{S} = \frac{1}{E_{\rm cut}} \,,
\end{equation}
where $E_{\rm cut}$ is the planewave energy cutoff required to achieve a desired accuracy with respect to  discretization in pseudopotential calculations.  Note that in defining the softness metric, we have set the softness of the all-electron Coloumb potential to be zero. Once the optimization problem has been solved for the Pareto frontier, we choose the following error metrics for selecting the pseudopotential (Eq.~\ref{Eq:Eq:MultiObjRefC}): $\Delta$-factor \cite{Delta2014}, lattice constant error $\delta_{\rm lat}$, error associated with the acoustic sum rule  in phonon calculations $\delta_{\rm asr}$, and error in phonon frequencies $\delta_{\omega}$ corresponding to a specified error in energy, having imposed the acoustic sum rule.  Indeed, these error metrics have been commonly used in literature to check the accuracy of pseudopotentials \cite{Delta2016reproducibility, PseudoDojo, SSSP}. Thereafter, the optimization problem for generating soft and transferable pseudopotentials can be written as:
\begin{subequations}
\begin{eqnarray}
\hat{\bf p} & =& \arg \min_{\mathbf{p} \in P} (\delta_{\Delta E}(\mathbf{p}), E_{\rm cut} (\mathbf{p})) \,,   \label{Eq:Eq:MultiObjRef1}\\ 
\mathbf{p}^*  & = &  \arg \min_{\mathbf{p} \in P_f} E_{\rm cut} (\mathbf{p}) \,, \quad P_f  =  \{\hat{\bf p}: \Delta(\hat{\bf p}) \leq \epsilon_1, \delta_{\rm lat}(\hat{\bf p}) \leq \epsilon_2, \delta_{\rm asr}(\hat{\bf p}) \leq   \epsilon_3, \delta_{\omega}(\hat{\bf p}) \leq \epsilon_4 \} \,, \label{Eq:Eq:MultiObjRefC1}
\end{eqnarray}
\end{subequations}
where the parameters that are being optimized $\mathbf{p}$ and the feasible set $P$ are dependent on the type of pseudopotential under consideration. 

	\section{ONCV pseudopotential formalism} \label{oncvpsp}

We now briefly review the ONCV pseudopotential formalism for completeness, the details of which can be found in Refs.~\cite{ONCVPSP,ONCVPSP_Erratum}. The atomic  nonlocal ONCV  pseudopotential operator can be written as 	
	\begin{equation}
		V^{\ell}_{\mathrm{nl}}  = \sum^{P^\ell}_{i,j=1}\ket{\chi^\ell_i}\big((B^\ell)^{-1}\big)_{ij}\bra{\chi^\ell_j} = \sum_{i=1}^{P^\ell}\ket{\tilde{\chi}^\ell_i}\frac{1}{\tilde{b}^\ell_i}\bra{\tilde{\chi}^\ell_i}, \quad B^\ell_{ij}=\braket{\varphi^\ell_i}{\chi^\ell_j}, i,j=1,\dots,P^\ell, \quad \ell=0,\dots,L, \label{vnl}
	\end{equation}
	where 
	$\varphi^\ell_i$ are the radial pseudo-wavefunctions,  $L$ is the maximum angular momentum, and 
	\begin{equation} \label{projectorChi}
		\ket{\chi^\ell_{i}} = (\varepsilon^\ell_i-T^\ell-V_{\mathrm{loc}})\ket{\varphi^\ell_i}, \quad i=1,\dots,P^\ell \,, 
	\end{equation}
	are the projectors obtained by $\varphi^\ell_i$ at energy $\varepsilon^\ell_i$. In addition, 	$\tilde{b}^\ell_i$ are the eigenvalues of $B^\ell$ and $\tilde{\chi}^\ell_i$ are linear combinations of  $\chi^\ell_i$ with coefficients derived from the eigenvectors, after having normalized $\chi^\ell_i$ and appropriately rescaled $B^\ell$. In Eq.~\ref{projectorChi}, $T^\ell=\left[ -d^2/dr^2 + \ell(\ell+1)/r^2\right]\!/2$ is the kinetic energy operator, and $V_{\mathrm{loc}}$ is the local potential operator that smoothly joins to the all-electron potential $V_{\mathrm{ae}}$ at radius $r_{\mathrm{loc}}$,  being arbitrary otherwise.
Let $\psi^\ell_{i}$ be the all-electron wave function corresponding to $\varphi^\ell_i$. Note that $\varphi^\ell_i(r)=\psi^\ell_i(r)$ for $r > r^\ell_c, \ell=0,\dots,L$ and $r_{\mathrm{loc}}\leqslant\min_{\ell=0,\dots,L}r^\ell_c$, where $r_{c}^\ell$ is the cutoff radius for angular momentum channel $\ell$.  The pseudo-wavefunctions are required to satisfy $M^\ell$ continuity constraints at $r^\ell_c$: 
	\begin{equation}\label{continuity}
		\left.\frac{d^m\varphi^\ell_i}{dr^m}\right\vert_{r^\ell_c} = \left.\frac{d^m\psi^\ell_i}{dr^m}\right\vert_{r^\ell_c},~ \quad m=0,\dots,M^\ell-1,~ i=1,\dots,P^\ell,~ \ell=0,\dots,L,
	\end{equation}
	as well as the generalized norm-conservation constraint:
	\begin{equation}\label{norm_const}
		\braket{\varphi^\ell_i}{\varphi^\ell_j}_{r^\ell_c}=\braket{\psi^\ell_i}{\psi^\ell_j}_{r^\ell_c},~\quad  i,j=1,\dots,P^\ell,~\ell=1,\dots,L \,,
	\end{equation}
	where the subscripts $r^\ell_c$ indicates that the domain of calculation is $[0,r^\ell_c]$. Note that  $B^\ell$ and therefore $V^\ell_{\mathrm{nl}}$ are Hermitian if the norm-conservation constraint (Eq.~\ref{norm_const}) is satisfied \cite{USPPs}. 
	
The pseudo-wavefunctions arising in the above expressions are determined as follows. For a given $\ell$, one can write the spherical Fourier transform of $\varphi^\ell_{i}(r)$ as $\bar{\varphi}^\ell_{i}(q) = \int_{0}^{\infty} j_\ell(qr)\varphi^\ell_{i}(r) r^2 dr$, where $j_\ell$ is a spherical Bessel function. Then, by considering a cutoff wavevector $q^\ell_c$, one can define the following residual kinetic energies:
	\begin{equation}\label{KE}
		E^{\ell}_{ij}(q^\ell_c) = \int_{q^\ell_{c}}^{\infty} \bar{\varphi}^\ell_{i}(q)\bar{\varphi}^\ell_{j}(q) q^4 dq 
		\equiv \enr{\bar{\varphi}^\ell_{i}}{\hat{E}^{\ell}(q^\ell_c)}{\bar{\varphi}^\ell_{j}}, \quad i,j=1,\dots,P^\ell,~\ell=0,\dots,L.
	\end{equation}
	Thereafter, the optimization problem for the pseudo-wavefunctions can be stated as follows:  given $\ell$, $q^\ell_c$, $r^\ell_c$, $M^\ell$, and $P^\ell$, find a set of pseudo-wavefunctions $\varphi^\ell_i, i=1,\dots,P^\ell$, that minimizes the resulting kinetic energies (Eq.~\ref{KE}) subject to continuity constraints (Eq.~\ref{continuity}) and norm-conservation constraint (Eq.~\ref{norm_const}).  
	To satisfy the continuity constraints (Eq.~\ref{continuity}), one first constructs a set of $N^\ell$ orthonormalized spherical Bessel functions $\xi_n^{\mathrm{O}\ell}$ such that $\xi_n^{\mathrm{O}\ell}=0$ for $r>r^\ell_c$, and writes $\varphi^\ell=\sum_{n=1}^{N^\ell} z^\ell_n\xi_n^{\mathrm{O\ell}}$. For simplicity, the subscript $i$ of $\varphi_i^\ell$ is ignored as this applies to all pseudo-wavefunctions with $i=1,\dots,P^\ell$. 
	One then substitutes $\varphi^\ell$ into Eq.~\ref{continuity} and uses singular value decomposition to solve the resulting system of $M^\ell$ linear equations in terms of $N^\ell$ unknowns $z^\ell_i, i=1,\dots,N^\ell$. The solution gives $\varphi^\ell_0 = \sum_{n=1}^{N^\ell} z^\ell_{0n}\xi_n^{\mathrm{O}\ell}, r \leqslant r^\ell_c; ~\varphi^\ell_0 =\psi^\ell, r > r^\ell_c$ and the basis functions $\xi_n^{\mathrm{N}\ell}, n=1,\dots,N^\ell-M^\ell$, spanning the null space of the linear system provided that $N^\ell>M^\ell$. Note that $\xi_n^{\mathrm{N}\ell}, n=1,\dots,N^\ell-M^\ell$, are orthonormal, orthogonal to $\varphi^\ell_0$, and have $M^\ell-1$ derivatives at $r^\ell_c$. 
	To further simplify the optimization problem, one can calculate the eigenvalues $e^\ell_n$ and eigenvectors of the matrix generated by $\enr{\xi_n^{\mathrm{N}\ell}}{\hat{E}^{\ell}}{\xi_s^{\mathrm{N}\ell}}$, and then define a set of ``residual'' basis functions $\xi_n^{\mathrm{R}\ell}$ as linear combinations of $\xi_n^{\mathrm{N}\ell}$ with coefficients derived from the eigenvectors such that $\enr{\xi_n^{\mathrm{R}\ell}}{\hat{E}^{\ell}}{\xi_s^{\mathrm{R}\ell}} = e^\ell_n\delta_{ns}, n,s=1,\dots,N^\ell-M^\ell$.	
	Now, by satisfying Eq.~\ref{continuity} and modifying Eq.~\ref{norm_const} and Eq.~\ref{KE} with the general solution:
	\begin{equation} 
		\varphi^\ell = \varphi^\ell_0 + \sum_{n=1}^{N^\ell-M^\ell} x^\ell_n \xi^{\mathrm{R}\ell}_n,
	\end{equation}
the optimization problem for finding a pseudo-wavefunction $\varphi^\ell$ can be simplified to read: given $\ell$, $q^\ell_c$, $r^\ell_c$, $M^\ell$, and $N^\ell>M^\ell$, find $x^\ell_n, n=1,\dots,N^\ell-M^\ell$,  such that 
	\begin{equation}
		E^{\ell} = E^{\ell}_{00} + \sum_{n=1}^{N^\ell-M^\ell}(2f^\ell_n x^\ell_n + e^\ell_n (x^\ell_n)^2),
	\end{equation}
	where $E^{\ell}_{00}=\enr{\varphi^\ell_0}{\hat{E}^{\ell}}{\varphi^\ell_0}$, and the ``force'' terms $f^\ell_n=\enr{\varphi^\ell_0}{\hat{E}^{\ell}}{\xi^{\mathrm{R}\ell}_n}$ are  minimized subject to norm-conservation constraint: 
	\begin{equation}
		\sum_{n=1}^{N^\ell-M^\ell}(x^\ell_n)^2 = \braket{\psi^\ell}{\psi^\ell}_{r_c}-\sum_{n=1}^{N^\ell}(z^\ell_{0n})^2 \equiv D^\ell_{\mathrm{norm}} .
	\end{equation}
Above, $D^\ell_{\mathrm{norm}}$ is the ``norm deficit'' of $\varphi^\ell_0$ relative to norm of $\psi^\ell$ on $\left[ 0, r^\ell_c \right]$.


\section{Implementation} \label{Sec:Implementation}
We have implemented the formulation described in Section~\ref{formulation} for the automated generation of soft and transferable ONCV pseudopotentials. In particular, we have developed a framework in \texttt{python}  that uses ONCVPSP \cite{ONCVPSP_code} for generation of the pseudopotentials, ABINIT \cite{ABINIT} for pseudopotential DFT calculations, and  Elk \cite{elk} for reference all-electron (i.e., Couloumb potential) DFT calculations. Given the large computational cost associated with Kohn-Sham DFT calculations, the framework is capable of running ABINIT as well as Elk in parallel,  allowing for significant reduction in the time to solution of the optimization problem described by Eqs.~\ref{Eq:Eq:MultiObjRef1} and \ref{Eq:Eq:MultiObjRefC1}. 

We choose the following ONCV pseudopotential parameters (Section~\ref{oncvpsp}):
\begin{equation}\label{Eq:ParametersONCV}
\mathbf{p} = (\{r^\ell_c\}_{\ell=0}^L,  \{q^\ell_c \}_{\ell=0}^L, \{N^\ell\}_{\ell=0}^L, \{M^\ell\}_{\ell=0}^L,  r_{\mathrm{loc}}) \,, 
\end{equation} 
where $\{r^\ell_c\}_{\ell=0}^L$ are the cutoff radii for the nonlocal projectors, $\{q^\ell_c \}_{\ell=0}^L$ are the cutoff wavevectors for the pseudo-wavefunctions' residual kinetic energies, $\{N^\ell\}_{\ell=0}^L$ are the number of basis functions used for the pseudo-wavefunctions, $\{M^\ell\}_{\ell=0}^L$ are the number of continuity constraints at $\{r^\ell_c\}_{\ell=0}^L$, and $r_{\mathrm{loc}}$ is the cutoff radius for the local part of the pseudopotential. The remaining  parameters in the ONCV formalism are held fixed at prespecified values, which in the current work are the same as in the PseudoDojo standard-accuracy PBE table \cite{PseudoDojo}. For the feasible set, we choose:
\begin{eqnarray}
P & = & \bigg \{ {\bf p}: \{r^\ell_c\}_{\ell=0}^L \in [1,r_u], \{q^\ell_c \}_{\ell=0}^L \in [1,14], \{N^\ell\}_{\ell=0}^L \in \{7,8,9\}, \nonumber \\ 
& & \{M^\ell\}_{\ell=0}^L = 4,  r_{\mathrm{loc}}\in\left[0.2,\min\nolimits \{r^{\ell}_c\}_{l=0}^L\right]  \bigg \} \,, \label{Eq:FeasibleSet}
\end{eqnarray}
which is generally decided on empirical considerations.

In the multi-objective optimization (Eq.~\ref{Eq:Eq:MultiObjRef1}), we choose $\delta_{\Delta E}$ to be the error metric corresponding to structural energy differences for the $14$ Bravais lattice symmetries (i.e., $m=1$), each with the nearest neighbor distance set to the equilibrium lattice constant for a simple cubic crystal, as determined from all-electron calculations using Elk. 
In isolated cases where the simple cubic crystal did not yield representative distances, the nearest neighbor distance was set instead based on known equilibrium distances. 
To reduce the computational cost of the optimization, we use a surrogate for $E_{\rm cut}$, namely $\tilde{E}_{\rm cut}$, which is defined to be the maximum cutoff energy over the different angular momentum channels (i.e., $\ell$) for kinetic energy convergence error of $10^{-5}$ Ha. This number is immediately available from the ONCVPSP code upon generation of the pseudopotential, and has been found to correlate well with the planewave energy cutoff $E_{\rm cut}$. Also, in order to eliminate pseudopotentials that have large errors in logarithmic derivatives and/or result in ghost states at lower energies, we augment the optimization problem with penalty functions within  the implementation. 

We solve the multi-objective optimization (Eq.~\ref{Eq:Eq:MultiObjRef1}) using the improved strength Pareto evolutionary algorithm (SPEA2) \cite{zitzler2001spea2}. The motivation for choosing an evolutionary algorithm is that (i) it does not require the calculation of derivatives of $\delta_{\Delta E}$ and $E_{\mathrm{cut}}$ with respect to $\mathbf{p}$, quantities that are not readily available in the current context, and (ii) the optimization problem is a mixed-integer non-linear problem with a computationally expensive black-box objective function.  During the optimization process, we choose the following settings in ABINIT and Elk:   Fermi-Dirac smearing of $0.001$ Ha, and $4\times4\times4$ Monkhorst-Pack grid for Brillouin zone integration with $(0.5, 0.5, 0.5)$ shift. In addition, we use planewave cutoff $E_{\mathrm{cut}}=60$ Ha in ABINIT and the highest accuracy \texttt{vhighq} option in Elk. Note that the all-electron Elk simulations need to be performed only once at the beginning of the optimization. In fact, storing the results allows for reuse in any subsequent efforts to generate pseudopotentials within the proposed formulation.

Once the Pareto frontier has been determined, we evaluate the  values of the error metrics $\Delta$, $\delta_{\rm lat}$, $\delta_{\rm asr}$, and $\delta_{\omega}$  for all the pseudopotentials belonging to this set. In particular, we calculate the $\Delta$-factor using the formalism/codes presented in Ref.~\cite{Delta2014} (with input files for ABINIT taken from Ref.~\cite{PseudoDojo}) which is set to  $\Delta$ (meV/atom); percentage error in lattice constants for structures used to test the GBRV pseudopotentials \cite{GBRV}, the mean of which is set to $\delta_{\rm lat}$ (\%); acoustic sum rule error corresponding to the highest and lowest acoustic frequencies (HAP and LAP, respectively) at the $\Gamma$-point   for the $\Delta$-factor lattice structures, the mean of which is set to  $\delta_{\rm asr}$ (cm$^{-1}$);  and  percentage phonon frequency error for the highest and lowest phonon frequencies (HOP and LOP, respectively, obtained after imposing the acoustic sum rule) at the $\Gamma$-point   for the $\Delta$-factor lattice structures, corresponding to an accuracy of $10^{-4}$ Ha/atom in the energy, the mean of which is set to $\delta_{\omega}$ (\%). For the ground state calculations, we choose the following settings in ABINIT and Elk: Fermi-Dirac smearing of $0.001$ Ha, and $8\times8\times8$  Monkhorst-Pack grid for Brillouin zone integration with $(0.5, 0.5, 0.5)$ shift. In addition, we use planewave cutoff of  $E_{\mathrm{cut}}=100$ Ha in ABINIT and the highest accuracy \texttt{vhighq} option in Elk. For the phonon calculations, we employ the density functional perturbation theory (DFPT) \cite{gonze1995adiabatic} feature in ABINIT. Note that since we have used a surrogate parameter for the planewave energy cutoff, we also determine the planewave energy cutoff in ABINIT corresponding to an accuracy in energy (with respect to discretization) of $\delta_E = 10^{-3}$ and $10^{-4}$ Ha/atom, as typical in practice, by choosing a simple cubic unit cell with equilibrium lattice constant determined by all-electron calculations,\footnote{Or by known equilibrium distances in isolated cases where the cubic system does not yeild representative distances.}  Fermi-Dirac smearing of $0.001$ Ha, and $8\times8\times8$  Monkhorst-Pack grid for Brillouin zone integration with $(0.5, 0.5, 0.5)$ shift. Once all these metrics have been calculated, the pseudopotential with the desired accuracy and maximal softness is selected.

	\section{Results and discussion}  \label{results}
We now use the aforedescribed formulation and implementation to generate a comprehensive table of soft and transferable ONCV pseudopotentials.  In particular, we consider scalar relativistic pseudopotentials with nonlinear core corrections in the Perdew–Burke–Ernzerhof (PBE) \cite{perdew1996generalized} exchange-correlation approximation. Within our framework, we use ONCVPSP 4.0.1, Elk 6.8.04, and ABINIT 8.10.3 for the construction of pseudopotentials, pseudopotential DFT calculations, and all-electron DFT calculations, respectively. The starting guess for the SPEA2 evolutionary algorithm, used to solve the multi-objective optimization (Eq.~\ref{Eq:Eq:MultiObjRef1}),  is a random set of  pseudopotentials that belong to the feasible set $P$ (Eq.~\ref{Eq:FeasibleSet}). The input parameters for the ONCVPSP code, other than those being optimized, are identical to those used to generate the standard-accuracy pseudopotentials in the PseudoDojo database \cite{PseudoDojo}. 

While in most cases highly accurate and soft potentials are generated directly, in some cases manual adjustments were required to improve transferability and/or smoothness. This can happen, for example, when the optimization is too aggressive in a given angular momentum channel, leading to sharp variations in projectors and/or insufficient log-derivative agreement with all-electron results. In most such cases, however, since the $\delta_{\Delta E}$ criterion targeted in the optimization encompasses a variety of crystal structures, the required adjustments were relatively straightforward, e.g., increasing $q_c^l$ in a given channel and/or $r_{\rm loc}$ of the local part. In the relatively few cases where adjustments were less straightforward, adjustment of individual projector $r_c^l$, local part $r_{\rm loc}$, and/or projector target energies was required to ensure desired smoothness and log-derivative agreement throughout the target spectrum.
	
\subsection{Example: magnesium}
\begin{figure}[htbp!]
		\centering
		\includegraphics[width = 0.93\textwidth]{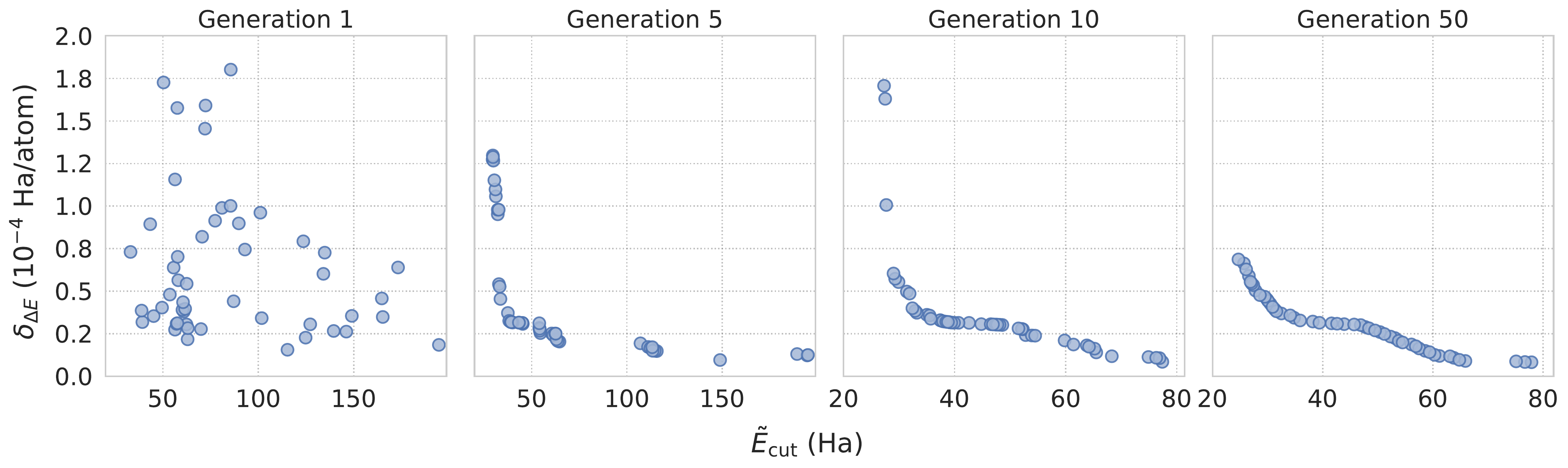} 
		\caption{Evolution of the Pareto frontier during the multi-objective optimization for the Mg pseudopotential.}
		\label{Fig:GAEvolution}
	\end{figure}

To clarify the construction process in practice, we consider Mg as a representative example.
In Fig.~\ref{Fig:GAEvolution}, we present the evolution of the Pareto frontier during the multi-objective optimization (Eq.~\ref{Eq:Eq:MultiObjRef1}). It is clear that around $50$ iterations/generations are sufficient to obtain a converged Pareto frontier, suggesting rapid convergence of the chosen evolutionary algorithm. Indeed, the number of iterations/generations required for convergence is dependent on the starting guess. The advantage of the developed formulation/implementation is that the pseudopotentials from the Pareto frontier can be used in subsequent efforts to generate optimized pseudopotentials, significantly accelerating convergence of the multi-objective optimization in such instances.

\begin{figure}[h!]
		\centering
		\includegraphics[width = 0.93\textwidth]{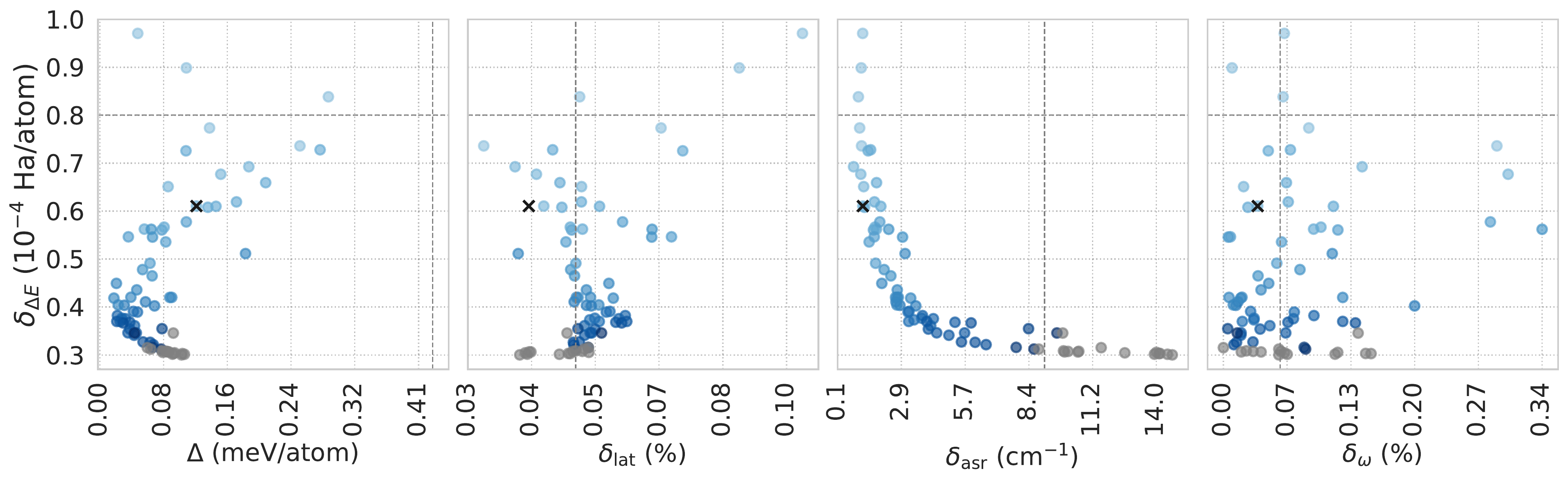} 
		\caption{Correlation between the error metric $\delta_{\Delta E}$ used in the multi-objective optimization and the error metrics $\Delta$, $\delta_{\rm lat}$, $\delta_{\rm asr}$, and $\delta_{\omega}$ used for the selection of the pseudopotential from the Pareto frontier, for magnesium. The dashed lines indicate the results for the PseudoDojo standard-accuracy pseudopotential. The shading used for the markers represents the softness of the pseudopotential, with lighter shades corresponding to softer  pseudopotentials. The gray color is used to represent pseudopotentials that are harder than PseudoDojo. The chosen pseudopotential is marked by the cross symbol.}
		\label{Fig:CorrelationErrMetrics}
	\end{figure}

In Fig.~\ref{Fig:CorrelationErrMetrics}, we plot the correlation between the error metric $\delta_{\Delta E}$ used in the multi-objective optimization (Eq.~\ref{Eq:Eq:MultiObjRef1}) and the error metrics $\Delta$, $\delta_{\rm lat}$, $\delta_{\rm asr}$, and $\delta_{\omega}$ used for the selection (Eq.~\ref{Eq:Eq:MultiObjRefC1}) of the pseudopotential from the Pareto frontier so generated. It is clear from the results that the chosen error metrics are not significantly correlated, providing motivation for their use in this work. In particular, as discussed before, $\delta_{\Delta E}$ is not designed to be a \emph{universal} metric for determining the accuracy of  pseudopotentials, but rather a computationally feasible one that can provide a good estimate of accuracy and transferability. Indeed, since $\Delta$, $\delta_{\rm lat}$, $\delta_{\rm asr}$, and $\delta_{\omega}$ are not significantly correlated with $\delta_{\Delta E}$, choosing all of them as part of the optimization problem is expected to produce pseudopotentials that are generally more transferable. It is also worth noting that more transferable pseudopotentials, as estimated by the error metrics considered, are not necessarily harder.  
This provides a key motivation for a systematic, optimization based approach.

\subsection{Table of ONCV pseudopotentials}	
\begin{figure}[h!]	
		\centering
		\includegraphics[width = 0.95\textwidth]{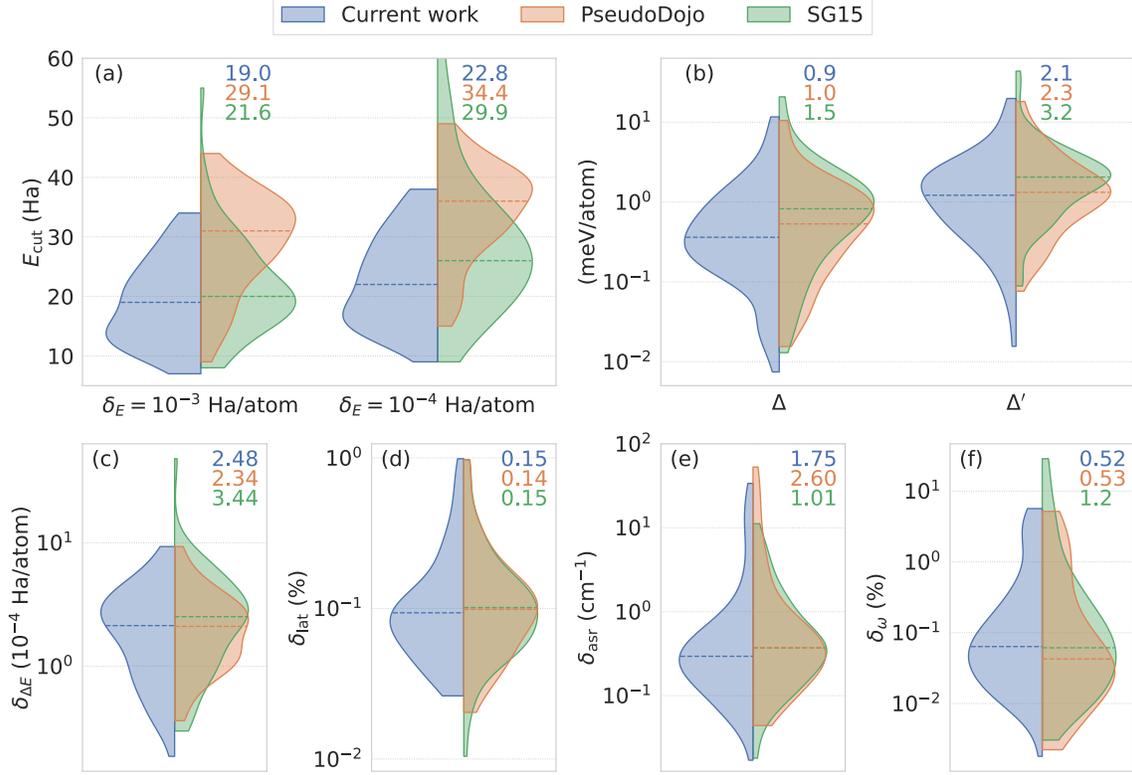} 
		\captionof{figure}{Overview of the softness and error metric values obtained for the present, SG15 \cite{SG15}, and  standard-accuracy PseudoDojo \cite{PseudoDojo} tables of  pseudopotentials.   The horizontal dashed lines correspond to median values and the numbers listed correspond to the mean values.
		} 
		\label{violin}
	\end{figure}

We now use the developed framework to generate a table of soft and transferable pseudopotentials for the sixty-nine chemical elements H--La and Hf--Bi (\url{https://github.com/SPARC-X/SPMS-psps}). The upper bounds of the error metrics, $\epsilon_1$, $\epsilon_2$, $\epsilon_3$, and $\epsilon_4$ (Eq.~\ref{Eq:Eq:MultiObjRefC1}) are chosen such that the present table has an accuracy comparable to that of the standard-accuracy PseudoDojo table \cite{PseudoDojo}. In Fig.~\ref{violin}, we present a summary of the softness and error metric values obtained for the present table, standard-accuracy PseudoDojo table \cite{PseudoDojo}, and SG15 table \cite{SG15} for comparison. Detailed results and comparisons for each chemical element are available at the SPMS-psps website. It is clear from the results that the present pseudopotentials have comparable accuracy to the PseudoDojo pseudopotentials, while being significantly softer. In particular, the average $E_{\rm cut}$ to achieve $\delta_E=10^{-3}$ Ha/atom is 18.7 Ha for the present table compared to 29.1 Ha for PseudoDojo. The corresponding numbers for $\delta_E=10^{-4}$ Ha/atom are 22.6 Ha and 34.4 Ha, respectively. These results translate to $\sim 2\times$ speedups in diagonalization-based DFT calculations having energy errors in the range  $10^{-3}$ to $10^{-4}$ Ha/atom as typical in practice. The speedups are significantly larger in the case of linear-scaling methods, particularly those that do not employ a reduced basis, e.g., speedups of $\sim 5\times$  are expected for the spectral quadrature method \cite{ suryanarayana2018sqdft, suryanarayana2013spectral}. It is also clear from the results that the present table of pseudopotentials is softer than the SG15 table while more accurate based on the metrics considered. Indeed, it is possible to choose $\epsilon_1$, $\epsilon_2$, $\epsilon_3$, and $\epsilon_4$ such that the present table of pseudopotentials has comparable accuracy to SG15, in which case the pseudopotentials generated could be made even softer.

	\section{Concluding remarks} \label{conclusions}

In this work, we employed multi-objective optimization to maximize pseudopotential softness while maintaining high accuracy and transferability.
We developed a formulation (Eqs.~\ref{Eq:Eq:MultiObjRef1} and \ref{Eq:Eq:MultiObjRefC1}) in which  softness and accuracy are simultaneously maximized, with accuracy $\delta_{\Delta E}$ (Eq.~\ref{Eq:Eq:deltadeltaE}) determined by the ability to reproduce all-electron energy differences between Bravais lattice structures, after which the resulting Pareto frontier is scanned for the softest pseudopotential that provides the desired accuracy in established transferability tests: $\Delta$-factor, lattice constant of GBRV structures, violation of acoustic sum rule in phonon calculations, and convergence of phonon frequencies. We employed an evolutionary algorithm to solve the multi-objective optimization problem and applied it to generate a table of ONCV pseudopotentials (\url{https://github.com/SPARC-X/SPMS-psps}) for the sixty-nine chemical elements H--La and Hf--Bi within the PBE exchange-correlation approximation.
We find that the resulting table is softer than the current PseudoDojo table of comparable accuracy and more accurate than the current SG15 table of comparable softness, according to the standard metrics considered.

Overall, the pseudopotentials generated using the proposed formulation/implementation are expected to significantly accelerate Kohn-Sham DFT calculations while maintaining accuracy comparable to existing high quality tables.  Indeed, initial versions of the pseudopotentials have already been successfully applied in the study of a number of low-dimensional systems \cite{bhardwaj2022elastic, bhardwaj2022strain, kumar2022bending}. The development of error metrics that provide a better quantification of accuracy and transferability, while remaining computationally tractable, is likely to further improve the quality and/or softness of the generated pseudopotentials, making it a worthy subject for future research. 
The generation of tables including spin-orbit coupling and additional exchange-correlation approximations will also be of interest to pursue.

	\section*{Acknowledgments}
	This work was supported by grant DE-SC0019410 funded by	the U.S. Department of Energy, Office of Science. This work was performed in part under the auspices of the U.S. Department of Energy by Lawrence Livermore National Laboratory under Contract DE-AC52-07NA27344. J.E.P gratefully acknowledges D.~R.~Hamann for invaluable discussions regarding the construction of ONCV pseudopotentials over the years. The views and conclusions contained in this document are those of the authors and should not be interpreted as representing the official policies, either expressed or implied, of the Department of Energy, or the U.S. Government.
	

%
%
%
%
%

	\bibliographystyle{unsrt}

\end{document}